\documentclass[useAMS,usenatbib]{mn2e}

\usepackage{indentfirst} 
\usepackage[dvips]{graphicx,color}
\usepackage{fancyhdr}
\usepackage{calc}
\usepackage{amssymb}
\usepackage{longtable}
\usepackage{times}
\usepackage{latexsym}

\title[GMRT observations of NGC 2997]
  {GMRT observations of NGC 2997 and radio\\
   detection of the circumnuclear ring}
\author[J. Kodilkar, N. G. Kantharia, S.~Ananthakrishnan] 
  {J.~Kodilkar,$^1$\thanks{Email: jitendra@ncra.tifr.res.in (JK); nimisha@ncra.tifr.res.in (NGK); subra.anan@gmail.com (SAK)}, N. G.~Kantharia,$^1$ S.~Ananthakrishnan $^2$\\
  $^1$National Centre for Radio Astrophysics, Tata Institute of Fundamental Research, Post Bag 3, Ganeshkhind, \\
  $^2$Dept. of Electronic Science, Pune University,\\ Pune 411 007, India}
\begin{document}

\date{Accepted 2011 May 12. Received 2010 Aug 09; in original form 2010 July 9}

\pagerange{\pageref{firstpage}--\pageref{lastpage}} \pubyear{2011}

\def\LaTeX{L\kern-.36em\raise.3ex\hbox{a}\kern-.15em
    T\kern-.1667em\lower.7ex\hbox{E}\kern-.125emX}

\newtheorem{theorem}{Theorem}[section]

\maketitle

\label{firstpage}

\begin {abstract}
We present high-resolution, high-sensitivity radio continuum observations of the
nearby spiral galaxy NGC 2997 at 332, 616 and 1272 $\rmn{MHz}$ using the Giant 
Metrewave Radio Telescope (GMRT). The integrated spectrum of this galaxy has a 
spectral index of $-0.92$ ($S_{\nu}\propto\nu^{\alpha}$) and we place an upper 
limit to the thermal fraction at 1272 $\rmn{MHz}$ of $\sim$ 10 per cent. Our 
multi-frequency study shows a relatively flat spectrum source ($\alpha\,\sim\,-0.6$) 
at the centre of the galaxy. This leads to radio detection of a circumnuclear ring 
in the high resolution map at 1272 $\rmn{MHz}$. We detect five hotspots in the ring, 
with average star formation rate of $\sim 0.024\,\rmn{M_{\odot}\,yr^{-1}}$, 
a median SN rate of $\sim\,0.001\,\rmn{yr^{-1}}$ and luminosity of $10^{20}\,\rmn{W\,Hz^{-1}}$. 
We estimate an equipartition field in the central nuclear region of diameter $\sim 750$ pc 
to be about $30\,\mu$G. We also report several interesting features along the spiral arms. 
In this paper, we present the low frequency radio continuum maps, 
the spectral index distribution, the circumnuclear ring and 
the derived physical properties.
\end{abstract}


\section{Introduction}
NGC 2997 is an interesting grand design spiral galaxy of type SAB(rs)c located 
in the loose galaxy group LGG 180 \citep{b14}. The disc inclination is 
$\sim 40^{\circ}$ \citep{b26} and the southern part of the disc is closer 
to us \citep{b29}. We assume a distance of 13.8 Mpc \citep{b33} 
to NGC 2997 and note that 1 arc-min corresponds to a linear scale of $\sim$ 4.0 kpc. 
Basic data on NGC 2997 are listed in Table \ref{tab1basic}.\\
\hspace*{6mm} The photometric and morphological analysis of this galaxy shows the presence
of a normal nucleus surrounded by several HII regions known as `hotspots' \citep{b28}.
Further investigations in optical \citep{b24}, ultraviolet \citep{b9} and
near-infrared \citep{b11} have detected a circumnuclear star-forming ring in NGC 2997. 
The study of such a ring is significant in terms of understanding gas transport to the 
inner regions of the galaxy and its association with any possible nuclear activity. 
NGC 2997 also shows several bright knots of emission along its arms. 
These knots are young star clusters with typical masses of $\sim 5\times 10^4$ M$_{\odot}$
\citep{b15}. This galaxy also hosts two giant HII regions which show supersonic 
velocity dispersion caused by an intense starburst \citep{b12}. 
The galaxy which is part of the HI brightest galaxy catalogue 
of the southern sky derived from HIPASS has an HI mass of $\sim 4.2\times10^9$ M$_{\odot}$ 
and half power line width of $254\,\rmn{km\,s^{-1}}$ \citep{b22}.\\
\hspace*{6mm}NGC 2997 has been extensively studied in radio wavelengths
shorter than $20$ cm by \citet{b17} and \citet{b25}. To the best of our knowledge, 
this galaxy has not been studied at longer wavelengths ($\lambda > 20$ cm). Since 
the non-thermal spectrum of synchrotron radiation is dominant at low frequencies, 
the GMRT observations at metre-wavelengths with high angular resolution and high 
sensitivity can reveal many significant features of NGC 2997 due to its moderate 
inclination such as smooth synchrotron emission from the disc, morphological structure 
of spiral arms, nucleus and compact HII regions. The spectral index distribution 
study of NGC 2997 could be useful for identification of giant HII regions.\\
\hspace*{6mm}The low radio frequency spectrum of the galaxy is important 
in obtaining the non-thermal spectral index of the galaxy which can then be used 
to separate the thermal and non-thermal contributions to the observed spectrum. 
Moreover, the non-thermal luminosities at low frequencies can be used to estimate 
the supernovae rate ($\rmn{yr^{-1}}$), average star formation rate ($\rmn{M_{\odot}\,yr^{-1}}$),
production rate of the Lyman continuum photons ($\rmn{s^{-1}}$) etc for the galaxy.\\
\hspace*{6mm}In this paper, we present radio continuum observation of NGC 2997 
using the GMRT \citep{b1,b31} at 332, 616 and 1272 $\rmn{MHz}$. Section 2 describes 
the observations and the data reduction procedure. Section 3 presents the total 
intensity maps and the morphological details of radio continuum features of NGC 2997. 
In section 4, we discuss the total emission spectrum of the galaxy, the spectral 
index map derived between 1272 and 332 $\rmn{MHz}$ and the circumnuclear ring. 
The last section summarizes the paper. 
\begin{table}\caption{Basic data on NGC 2997$^{\dag}$}\label{tab1basic}
\begin{tabular}{@{}lc}
\hline
Optical center (J2000)       &09h45m38.8s  -31d11$'28''$\\
Morphological type$^{\ddag}$ &     SAB(rs)c       \\
Major diameter (arcmin) &  8.9              \\
Minor diameter (arcmin) &  6.8              \\
Distance ($H_{0}=75\,\rmn{km\,s^{-1}\,Mpc^{-1}}$)   &13.8 Mpc  \\
Magnitude (Visual) &        10.06         \\
Inclination angle (deg.)     &       40             \\
Heliocentric radial velocity ($\rmn{km\,s^{-1}}$) & 1088 $\pm$ 2   \\
\hline
\end{tabular}
\dag Reference : NASA/IPAC Extragalactic Database (NED)\\
\ddag \citet{b44}
\end{table}
\section{Observations and data reduction} 
We had a total of three observing sessions on NGC 2997 using the GMRT synthesis 
array at each of the three frequency bands 332, 616 and 1272 $\rmn{MHz}$ 
respectively. Observing sessions were conducted in the period from 2003 to 2004 
using the standard spectral line mode of the GMRT digital correlator which gives 
visibility data of 16 $\rmn{MHz}$ bandwidth across 128 channels for two polarisations. 
The observational parameters are summarized in Table \ref{table2obs}. The raw visibility 
data were converted to FITS and analysed using standard AIPS\footnote{Astronomical Image Processing System, distributed by the National Radio Astronomy Observatory, http://www.aips.nrao.edu. The NRAO is a facility of the National Science Foundation operated under cooperative agreement by Associated Universities, Inc.}.\\
\hspace*{6mm}The VLA flux density calibrators, 3C147 and 3C286 were used as flux 
density reference to scale the flux densities of phase calibrators and the target 
source. To correct the ionospheric and instrumental gain variations, phase calibrator 
0837-198 was observed for 6 $\rmn{min}$ during every 25 $\rmn{min}$ of observation of 
the target source. The flux density values of phase calibrators obtained using the task 
GETJY are given in Table \ref{table2obs}. The flux density calibrators were also used 
for bandpass calibration. To avoid bandwidth smearing effect, the band-pass calibrated 
data at 616 and 1272 $\rmn{MHz}$ band were collapsed to five channels each of $\sim$2.75 $\rmn{MHz}$ 
bandwidth by averaging every 22 channels. At 332 $\rmn{MHz}$, ten channels each 
of $\sim$1.25 $\rmn{MHz}$ bandwidth formed from RFI free band by averaging every 10 channels.
The expected error bars in flux density scale for the GMRT observations are $\lesssim\,10$ per cent.\\
\begin{table*}
\begin{minipage}{140mm}
\caption{Observation table}
\label{table2obs}
\footnotesize
\begin{tabular}{@{}llccc@{}}
\hline
     & Observing band ($\rmn{MHz}$)      &  \textbf{332}   &  \textbf{616}   &\textbf{1272}\\
\hline
1\@. & Date                      &    2004 Feb 24& 2003 Jan 21  & 2003 Aug 16\\
2\@. & Correlator used (USB/LSB)$^{a}$ &      USB        &      USB        &       LSB   \\
3\@. & On source time(Hrs.)$^{b}$&      5          &      4          &         3  \\
4\@. & Receiver bandwidth ($\rmn{MHz}$)  &      16         &      16         &       16    \\
5\@. & No. of working antennas$^{c}$      &      27         &      24         &       28    \\
6\@. & Shortest spacing(k$\lambda$)&    0.06       &     0.120       &     0.250     \\
     &                           &     $\sim55\,\rmn{m}$   &     $\sim60\,\rmn{m}$   &   $\sim58\,\rmn{m}$ \\
7\@.& Longest spacing(k$\lambda$) &     27        &      40          &     101     \\
     &                           &     $\sim25\,\rmn{km}$   &     $\sim20\,\rmn{km}$  &   $\sim24\,\rmn{km}$ \\ 
8\@.& Largest visible structure$^{d}$ &  $\sim$19$'$   &     $\sim$12$'$             &    $\sim$8$'$  \\
9\@.& Flux density calibrator(s) &     3C147,3C286
 &  3C147,3C286   &   3C147,3C286\\
     & Flux density in Jy $^{e}$&     52.69,25.96 & 38.26,21.07     & 23.69,15.48  \\
10\@.& Phase calibrator(s)       & 0837-198,0902-142&  0837-198      &  0837-198   \\
     & Flux density in Jy $^{f}$&$13.3\pm0.4, 4.3\pm0.1$&$8.7\pm0.08$&$4.6\pm0.1$\\
\hline
\end{tabular}
\medskip
\\a. USB-Upper side band, LSB-Lower side band. b. Total observation time on the object before editing. c. Maximum number of antenna operational at any time during the observation. d. Corresponding to the shortest spacing present in our data. e. Set by SETJY task : Using VLA (1999.2) or Reynolds (1934-638) coefficients. f. Flux density and error from GETJY.
\end{minipage}
\end{table*}
\hspace*{6mm}First, a low resolution map of $\sim$ $45$ arcsec was made using `uvtaper'
at each observing frequency and the entire primary beam was imaged. We also used 
varying tapers to generate maps of different angular resolution which can be compared 
at the different frequencies. There were two to three strong point sources near the 
edge of primary beam at each frequency with peak flux ten to twelve times stronger 
than the peak flux on the object. These sources were removed using the task UVSUB.\\
\hspace*{6mm}Wide field imaging was used at 616 and 332 $\rmn{MHz}$. The number of 
facets used for imaging NGC 2997 at 616 and 332 $\rmn{MHz}$ were 13 and 22 respectively 
(obtained using SETFC task in AIPS). No self calibration was done at 1272 $\rmn{MHz}$. 
At other frequencies, the visibility data sets were first phase self-calibrated using 
only strong point sources near or within the inner quarter of the primary field. After 
two to three iterations of phase self-calibration using only point sources, the source 
model obtained by cleaning up to the rms noise level of the map was included in 
self-calibration. At 332 and 616 $\rmn{MHz}$, it took about four to six iterations to 
get good convergence of self-calibration resulting in improved dynamic range maps. 
All facets were combined and primary beam correction applied using the FLATN task
to produce a map at each observing frequency. 
\section{Observational results}

\subsection{Total intensity maps}

\begin{figure*}
\centering
\vbox to 220mm{\vfil
\caption{The total intensity maps of NGC 2997 with a resolution of 15 arcsec at 1272, 616, and 332 $\rmn{MHz}$.}\label{n2997fig1}
\begin{tabular}{@{}cc@{}}
\includegraphics[height=7cm,width=8cm]{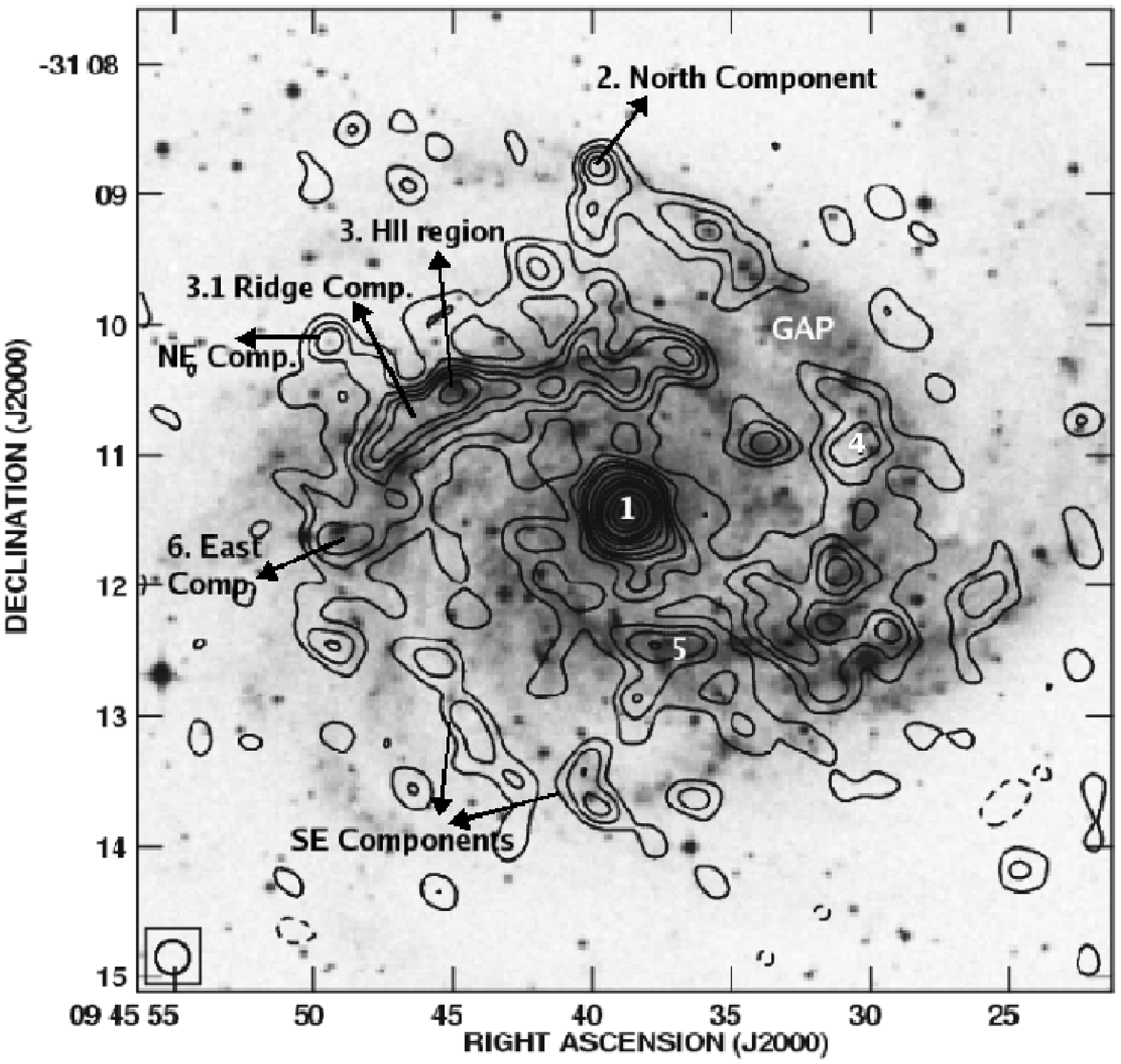}& \parbox[b][70mm]{60mm}{ \small{ \textbf{Fig. \ref{n2997fig1}a :}\footnotesize{ The 1272 $\rmn{MHz}$ radio contour map superimposed on the optical Digitized Sky Survey image. The six discrete radio continuum features detected at three wavebands are labelled with numbers. The peak brightness is 21 $\rmn{mJy\,beam^{-1}}$. The rms $\sigma$ is 0.3 mJy per beam area, the beam is depicted at the bottom left corner. The contours are at ( -3, 3, 5, 7, 9, 11, 15, 20, 25, 30, 40, 50, 60, 70 ) $\times~\sigma$ }}}
\\
\newline
\includegraphics[height=7cm,width=8cm]{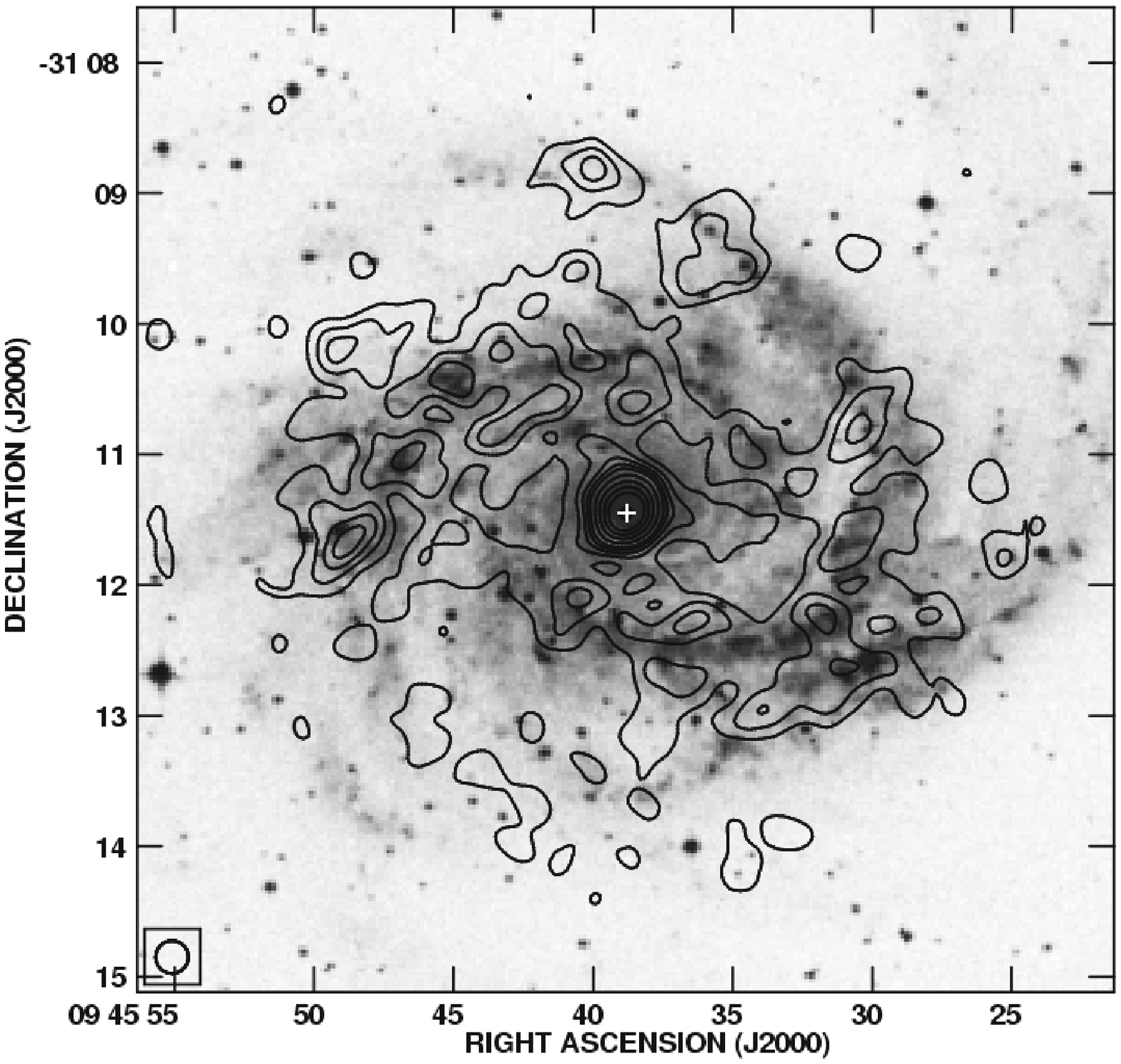}& \parbox[b][70mm]{60mm}{ \small{\textbf{Fig. \ref{n2997fig1}b :}\footnotesize{ The 616 $\rmn{MHz}$ radio contour map overlaid on the DSS optical image. The radio peak position of the nucleus is marked as a plus sign. The peak brightness is 28.4 $\rmn{mJy\,beam^{-1}}$. The rms $\sigma$ is 0.6 mJy per beam area, the beam size is depicted at the bottom left corner. Contours : ( -3, 3, 5, 7, 9, 11, 15, 20, 25, 30, 40 ) $\times~\sigma$ }}}
\\
\newline
\includegraphics[height=7cm,width=8cm]{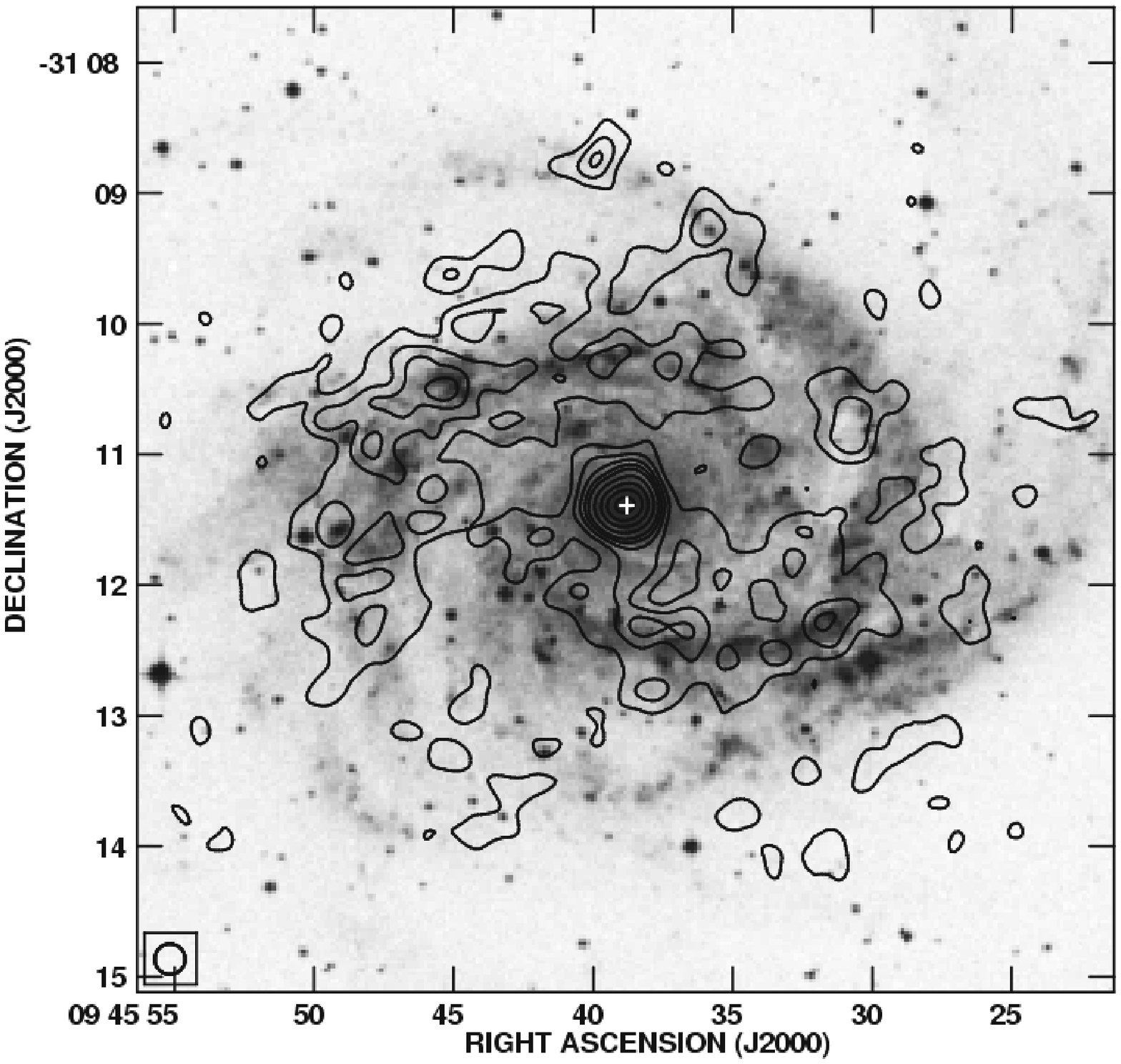} & \parbox[b][70mm]{60mm}{ \small{ \textbf{Fig. \ref{n2997fig1}c :}\footnotesize{ The 332 $\rmn{MHz}$ radio contour map overlaid on the DSS optical image. The radio peak position of the nucleus is marked as a plus sign. The peak brightness is 40.9 $\rmn{mJy\,beam^{-1}}$. The rms $\sigma$ is 1 mJy per beam area, the beam size is shown at the BLC. Contours : ( -3, 3, 5, 7, 9, 11, 15, 20, 25, 30, 40 ) $\times~\sigma$.}}}
\end{tabular}
\vfil}
\end{figure*}

We made total intensity maps of the object with angular resolution ranging
from 45 to 3 arcsec. Whenever possible, the total intensity maps at different 
observing frequencies were made with a similar angular resolution by using the 
uv-tapering function in the task `IMAGR'. Fig.\ref{n2997fig1} shows the total 
intensity maps of NGC 2997 at 1272, 616 and 332 $\rmn{MHz}$ at a resolution of 15 arcsec 
overlaid on the optical Digitized Sky Survey image. The uv-tapering and weighting functions 
used while mapping the object are given in Table \ref{n2997tab0}.\\
\begin{table*}
\begin{minipage}{140mm}
\caption{Map parameters}\label{n2997tab0}
\footnotesize
\begin{tabular}{@{}|l|c|c|c|@{}}
\hline
\textbf{ Figure }             &  \textbf{\ref{n2997fig1}a}       & \textbf{\ref{n2997fig1}b}    & \textbf{\ref{n2997fig1}c} \\
\hspace{5mm}Band Center ($\rmn{MHz}$)\dag       & \textbf{1272}           &  \textbf{616}       & \textbf{332}  \\
\hspace{5mm}Synthesized Beam$^a$& $15.8''\times12.4''$ @$25^{\circ}$&$17.9 ''\times11.7''$@$41^{\circ}$ &$15.9''\times12.0''$@$14^{\circ}$\\
\hspace{5mm}Restoring Beam$^b$ &         $15''$         &    $15''$      &       $15''$       \\
\hspace{5mm}UV-taper/Weighting      &  $ 20\,\rmn{k\lambda}$/NA&    $ 20\,\rmn{k\lambda}$/NA &  $ 16\,\rmn{k\lambda}$/UN \\
\hspace{5mm}RMS noise ($\rmn{mJy\,beam^{-1}}$)  &          0.3        &         0.6         &        1.0     \\
\hline
\textbf{ Figure}   &    \textbf{\ref{n2997fig2}}         &  \textbf{\ref{circumn2997fig}-contour}        &  \textbf{\ref{circumn2997fig}}-left grey$\,^c$ \\
\hspace{5mm}Band Center ($\rmn{GHz}$)\dag  & \textbf{1.27}   & \textbf{1.27}      &  \textbf{4.8}       \\ 
\hspace{5mm}Synthesized Beam$^a$ &$50.9''\times36''$@$11^{\circ}$&$4.06''\times2''$@$11^{\circ}$&$1.5''\times0.7''$ @$55^{\circ}$  \\ 
\hspace{5mm}Restoring Beam$^b$   &    $45''$   &       $3''$           &      $  3''$        \\ 
\hspace{5mm}UV-taper/Weighting   &$ 5\,\rmn{k\lambda}$/NA &  None/UN  & None/UN  \\ 
\hspace{5mm}RMS noise ($\rmn{mJy\,beam^{-1}}$) &           0.6          &          0.08        &         0.04         \\ 
\hline
\end{tabular}
\medskip
\newline\dag All maps are stoke I and corrected for the shape of the GMRT primary beam.\\
\hspace{1cm} NA-Natural weighting, UN-Uniform weighting.\\
\hspace{1cm} Final bandwidth for analysis after editing is about $\sim 12.5$ $\rmn{MHz}$ at each frequency.\\
a. Synthesized beam for a deconvolution, position angle with respect to the major axis is given in degree.\\
b. Restoring beam is a circular Gaussian with PA=0 degree, and the area is approximated to the 
\\~devoncolving beam area in square arcsec $\Omega_{s}=1.13 * \beta_{maj}\times\beta_{min}$.\\
c. Map using the VLA archival data.
\end{minipage}
\end{table*}
\normalsize 
\hspace*{6mm} The multifrequency total intensity GMRT maps in Fig. \ref{n2997fig1}
show about six common radio continuum features of NGC 2997. These discrete features 
are labelled by numbered tags in Fig.\ref{n2997fig1}a. In Table \ref{n2997dtab}, we summarize 
their positions, size, observed flux densities, etc. The peak positions of the discrete features 
listed in Table \ref{n2997dtab} match within median errors of $\Delta\rmn{RA}\,\sim \pm0.22^{s}$ 
and $\Delta\rmn{DEC}\,\sim \pm 2$ arcsec. The summary of radio continuum emission features seen in 
NGC 2997 is as follows :\\

\begin{figure}
\includegraphics[height=7cm,width=8cm]{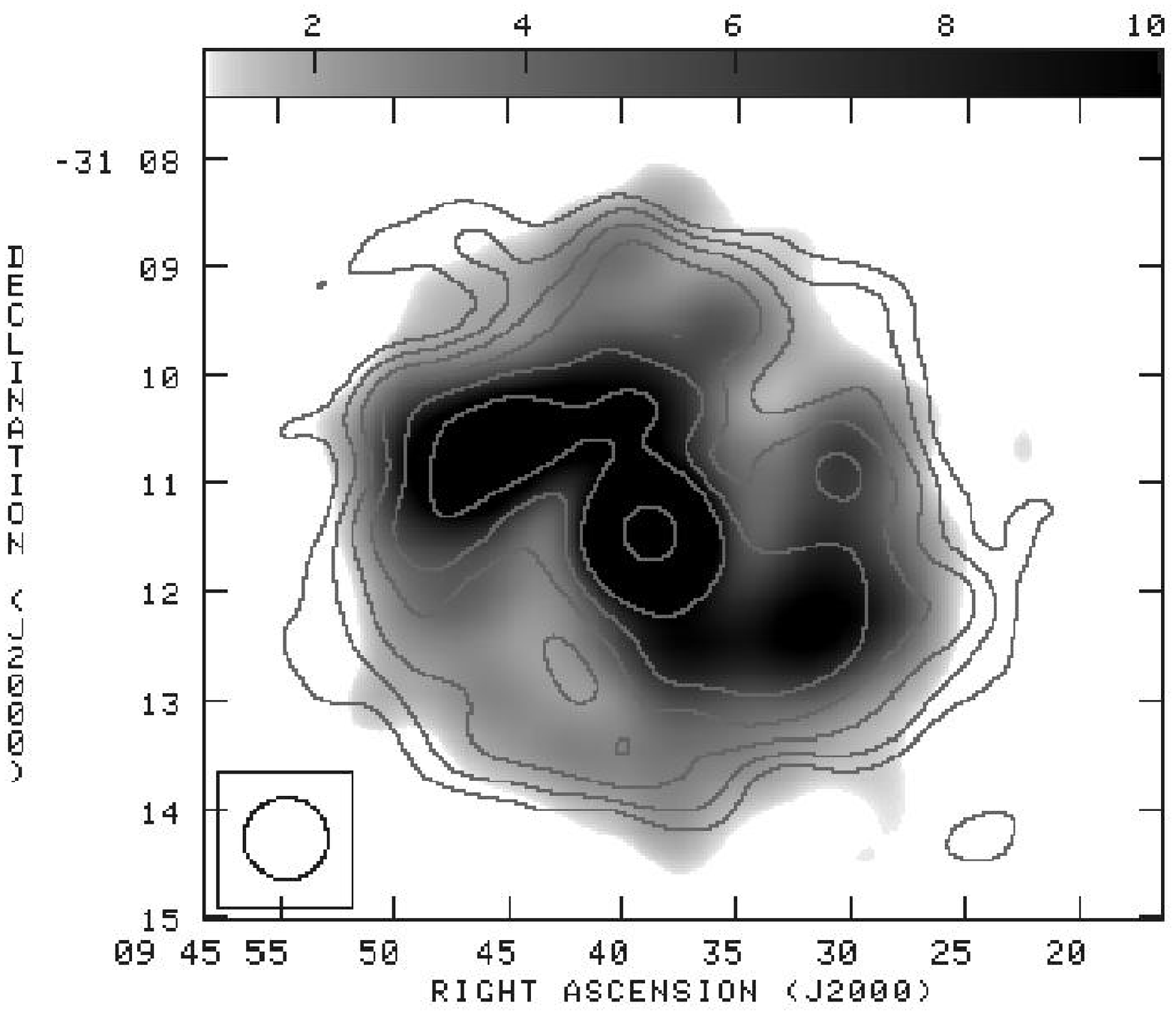}
\caption{shows the GMRT 1272 $\rmn{MHz}$ radio contour map of NGC 2997 with a beamsize of 45 arcsec overlaid on the NVSS 1400 $\rmn{MHz}$ grey scale image \citep{b7}.  The disc extent and morphology of the GMRT low resolution map is similar to the NVSS image indicating that the GMRT map is sensitive to large scales. The rms $\sigma$ is 0.6 mJy per beam area, contours starting from $3\,\sigma$ level and scaled by factor $\sqrt{2}$.}\label{n2997fig2}
\end{figure}

\textbf{\small{(i)}}\normalsize~The radio disc in the GMRT
 low-resolution images has an extent 
of $\sim$ $6.5'$ by $8'$ which is comparable to the optical extent. The extent of radio 
emission in the low resolution map at 1272 $\rmn{MHz}$ is comparable to the NVSS grey-scale 
image \citep{b7} as shown in Fig.\ref{n2997fig2}.\\
\textbf{\small{(ii)}}\normalsize~All our total intensity maps show the compact bright 
nucleus, centred at, RA$=09^{h}45^{m}38.76^{s}$; DEC$=-31^{\circ}11'26.60''$ (J2000). 
The uncertainties in RA and DEC are $\sim\,0.03^s$ and $\sim\,0.4$ arcsec, respectively. 
The radio peak position coincides with the values given by \citet{b17}, 
\citet{b25} and also the optical centre (see Table \ref{tab1basic}). 
At first glance, the intense unresolved nucleus at 15 arcsec resolution (Fig.\ref{n2997fig1})
appears like an AGN.\\
\begin{table*}
\centering
\begin{minipage}{144mm}
\caption{Some discrete components in the disc of NGC 2997}\label{n2997dtab}
\scriptsize
\begin{tabular}{@{}lcccccl@{}}
\hline \\[-2.0ex]
Source &  Observed position &\multicolumn{3}{c}{Integrated flux densities ($\rmn{mJy}$)} & Spectral index  & Comments \\
       &  at 1272 $\rmn{MHz}$       &\multicolumn{3}{c}{\& Peak intensities ($\rmn{mJy\,beam^{-1}}$) at} &  332-1272 $\rmn{MHz}$ & \\[0.5ex]
       &$\alpha_{2000}=\rmn{h\,m\,s\,\,\pm\,s}$&\cline{1-3}& & \\ [-2.0ex]
       &$\delta_{2000}=\,\rmn{^{\circ}\,\,\,'\,\,''\,\,\,\pm''}$  &  1272 $\rmn{MHz}$ & 616 $\rmn{MHz}$  &  332 $\rmn{MHz}$ &($S_{\nu}\propto\nu^{\alpha}$) &  \\ [0.8ex]
\hline\\ [-1.5ex]
 1   & 09 45 38.76 $\pm$0.0 & 38.7 $\pm$1.6 & 56.6 $\pm$3.1 &86.3$\pm$4.0 & -0.6 & Nuclear region        \\
     &-31 11 26.60 $\pm$0.3 & 21.0 $\pm$0.5 & 28.4 $\pm$1.2 &40.9 $\pm$1.3 &-0.5 &                      \\[1.2ex]
 2   & 09 45 39.65 $\pm$0.1 & 13.7 $\pm$2.4 & 20.9 $\pm$4.8 &21.0 $\pm$4.6 &-0.31 & North source           \\
     &-31 08 51.15 $\pm$2.1 & 4.2  $\pm$0.6 & 6.9  $\pm$1.2 &8.1  $\pm$1.3 &-0.48 &                        \\[1.2ex]
 3   & 09 45 45.05 $\pm$0.1 & 16.7 $\pm$2.1 & 14.0 $\pm$4.6 &12.04$\pm$3.6 & 0.24 & HII region in northern\\
     &-31 10 33.08 $\pm$1.1 &  6.0 $\pm$0.6&  4.9  $\pm$1.2 & 6.5 $\pm$1.4 &-0.06 & arm                   \\[1.2ex]
 3.1 & 09 45 45.87 $\pm$0.3 & 19.3 $\pm$3.7 & 27.0 $\pm$5.2 & 59.8$\pm$14&  -0.84   & Northern ridge\\
     &-31 10 41.19 $\pm$2.3 &  6.3 $\pm$0.5&  6.8  $\pm$0.9 & 15.2 $\pm$1.2&       & component ($\sim$5.0 $\rmn{kpc}$ by 2.1 $\rmn{kpc}$)\\
 4   & 09 45 30.53 $\pm$0.2 & 5.3 $\pm$2.0 & 18.3 $\pm$5.6 & 22.6$\pm$4.9 & -1.08 & Nonthermal north-west  \\
     &-31 10 54.90 $\pm$4.1 & 2.1  $\pm$0.6 & 5.0  $\pm$1.2 & 8.0 $\pm$1.3 &-0.98 & component            \\[1.2ex]
 5   & 09 45 37.42 $\pm$0.3 & 6.1 $\pm$2.1 &  5.8 $\pm$3.2 & 15.6$\pm$5.0 & -0.69 & Component in southern arm\\
     &-31 12 28.48 $\pm$2.3 & 2.25  $\pm$0.5 & 3.42  $\pm$1.3 & 5.4 $\pm$1.3& -0.65 &            \\[1.2ex]
 6   & 09 45 48.84 $\pm$0.2 & 26.1 $\pm$4.4 & 13.3 $\pm$4.1 & 22.5$\pm$5.3 & 0.11 &East component of \\
     &-31 11 36.10 $\pm$2.5 & 3.9  $\pm$0.5 & 5.4 $\pm$1.2 & 7.3 $\pm$1.3 & -0.45 &North-east arm    \\[1.2ex]
\hline
\end{tabular}
\end{minipage}
\end{table*}
\textbf{\small{(iii)}}\normalsize~The radio continuum emission associated with both 
the spiral arms is clearly detected in all the maps shown in Fig.\ref{n2997fig1}. 
There is an intense ridge of emission along the northern spiral arm as compared to 
the southern arm.\\
\textbf{\small{(iv)}}\normalsize~No radio emission is detected in the northern part 
of the southern spiral arm near the region at RA=9$^{h}$45$^{m}$32.8$^{s}$; DEC$=-31^{\circ}10'$
(labelled as `GAP' in Fig. \ref{n2997fig1}a). This region is bluer in the optical image, 
suggesting a young star forming complex. This gap has also been noted in the VLA total 
intensity maps at $\lambda20$, 18, $6$ cm and in the polarized intensity maps 
at $\lambda6$ and $13\,$cm \citep{b17,b25}.\\
\textbf{\small{(v)}}\normalsize~A bright compact source (component 2) is seen in all
our maps of NGC 2997 at RA$=09^{h}45^{m}39.6^{s}$; DEC$=-31^{\circ}08'51.1''$,
in the north of the galaxy. No discrete optical counterpart is seen for this source.\\
\textbf{\small{(vi)}}\normalsize~All the maps in Fig.\ref{n2997fig1} show 
an intense region (component 3) in the northern spiral arm, near 
RA$=09^{h}45^{m}45^{s}$; DEC$=-31^{\circ}10'33''$. 
This compact region is coincident with the giant HII region identified by 
\citet{b12} using high resolution optical spectra ($\lambda3726-7136\,\rmn{\AA}$).
Apart from this, a prominent north-east discrete component (labelled as `NE Comp.') 
is seen in Figs.\ref{n2997fig1}a and \ref{n2997fig1}b, positioned at RA=09$^h$45$^m$49.28$^s$, DEC$=-31^{\circ}10'11.02''$ 
(J2000) and lying roughly above the northern ridge. This component appears to be a flat spectrum 
source since it is barely detected in the 332 $\rmn{MHz}$ map (Fig. \ref{n2997fig1}c).\\
\small{\textbf{Global radio continuum properties :}}\\\normalsize
The integrated flux densities given in Table \ref{n2997gb} are obtained by 
integrating the region above $3\sigma$ contour in the low resolution 
maps (45 arcsec). The measured spectral power ranges from 
$\sim 7.9\,\times\,10^{21}\,\rmn{W\,Hz^{-1}}$ at 1272 MHz to $24.4\,\times\,10^{21}\,\rmn{W\,Hz^{-1}}$ 
at 332 MHz. At 1.4 $\rmn{GHz}$, normal galaxies range in power from L
$\lesssim 10^{18}\,\rmn{to} \sim10^{23}\,\rmn{h^{-2}\,W\,Hz^{-1}}$ where
h$\equiv H_{0}/(100\,\rmn{km}\,\rmn{s}^{-1}\,\rmn{Mpc}^{-1})$ \citep{b4}.
The global spectral index values given in Table \ref{n2997gb} indicate 
that the spectrum flattens at the lower frequencies. The total emission spectrum 
of many normal spiral galaxies with moderate surface brightness are known to 
show a break in the range from 0.1 to 1 $\rmn{GHz}$. One of the reasons 
for this is possibly free-free absorption by the cool ($T_{e} < 1000\,\rmn{K}$) 
ionized gas filling a large fraction of the radio emitting volume \citep{b20}.
On the other hand, the break could be due to propagation effects of relativistic
electrons and subsequent energy losses \citep{b38,b39}\\
\begin{table*}
\centering
\begin{minipage}{140mm}
\caption{Global radio continuum properties of NGC 2997}\label{n2997gb}
\footnotesize
\vspace{0.3cm}
\begin{tabular}{@{}lccc@{}}
\hline
Observing band ($\rmn{MHz}$) $\longrightarrow$ &  \textbf{332}   &  \textbf{616}   &\textbf{1272}\\
\hline
    &             &             &              \\
Flux density (mJy)$\,^{\dag}$& 1134 $\pm111$ & 730 $\pm22$ & 367 $\pm8$ \\
Spectral Power ($10^{21}\,\rmn{W\,Hz^{-1}}$)& 24.4 $\pm1.7$ & 15.7 $\pm0.3$& 7.9 $\pm0.1$\\
    &             &             &              \\
Global Spectral Index ($S_{\nu}\propto\nu^{\alpha}$) &$332\rightarrow 616 : -0.71$ &$616\rightarrow 1272 : -0.94$&$332\rightarrow 1272 : -0.84$\\
\hline
\end{tabular}\\
\dag The reasonable error bars in flux density scale are expected to be $\lesssim\,10$ per cent.
\end{minipage}
\end{table*}\normalsize
\hspace*{6mm}The radio-selected samples of normal galaxies are characterized 
by the FIR/radio flux density ratios \citep{b4}. \citet{b18}
defined the parameter $q$ as a logarithmic measure of the FIR/radio flux 
density ratio. For more general comparisons, the $\hat{q}_{\rmn{FIR}}$ is defined as 
$\hat{q}_{\rmn{FIR}}=\rmn{log\{S_{\nu}(FIR)/[S_{\nu}(rad)}\times(\rmn{\nu_{rad}/1.4\,\rmn{GHz}})^{-\alpha})]\}$ 
\citep{b13}, where $\alpha$ is the radio spectral index ($S_{\nu}\propto\nu^{\alpha}$). 
The spectral index for the observed GMRT data is -0.79 $\pm$0.07. 
From the {\it IRAS\/} surveys \citep{b13} at 60 and 100 $\rmn{\mu\,m}$ 
the flux densities
 of NGC 2997 are 32.28 and 85.14 Jy, respectively. 
Using the {\it IRAS\/} FIR flux density measurements and integrated flux densities 
measured with the GMRT, the derived $\hat{q}_{\rmn{FIR}}$ parameter 
is $\sim 2.19 \pm0.03$, which is close to the median for spiral galaxies at 1.4 $\rmn{GHz}$,
i.e. $\hat{q}_{\rmn{FIR}}\approx 2.30$ with the rms scatter $\sigma_{\hat{q}}\lesssim 0.2$ 
\citep{b5,b8,b18}. The spectral power and FIR/radio flux density ratio found at our 
observing frequencies confirms that NGC 2997 is a normal spiral galaxy.
\section{Discussion}

\subsection{The observed radio spectrum \& SFR}

\begin{figure}
\includegraphics[height=8cm,width=5cm,angle=-90]{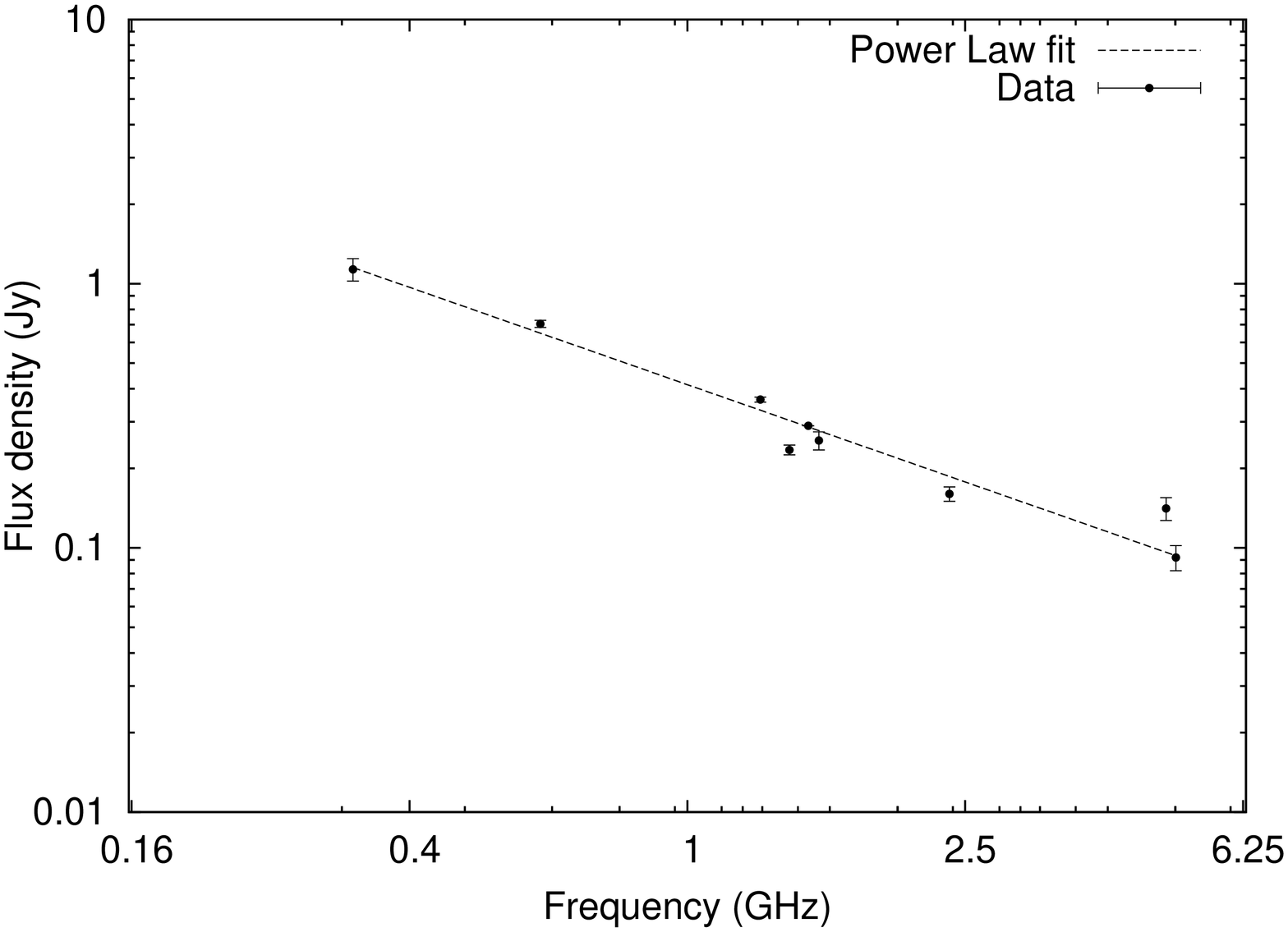}\\
\caption{The total emission spectra of NGC 2997. In addition to our three
GMRT data points, the data given in Table \ref{tp_tab} are also plotted. The dashed
line is the best least-squares fitting to the data with a power law index $\alpha =-0.92\,\pm$0.04}\label{tpcurv_fig}
\end{figure}

Integrated flux densities of NGC 2997 obtained using the GMRT and other 
published data from 0.33 to 8.46 $\rmn{GHz}$ are listed in Table \ref{tp_tab}. 
A single power law ($S_{\nu}\propto \nu^{\alpha}$) with $\alpha=-0.92\,\pm$0.04 gives 
the best-fitting to the radio spectrum from 0.33 to 5 $\rmn{GHz}$ (Fig. \ref{tpcurv_fig}). 
A value of $\alpha=-1.10\,\pm0.07$ is quoted by \citet{b25} for the radio spectrum from 
1.43 to 8.46 $\rmn{GHz}$. \citet{b45} analysed the radio continuum spectra of 74 galaxies 
to separate the thermal and the non-thermal radio emission. The mean value obtained for non-thermal 
spectral index is $\langle\alpha_{nt}\rangle=-0.83\pm0.02$. They also studied the correlation
of the morphological type and non-thermal spectral index and found that spiral galaxies of type Sc 
show $\alpha_{nt}$ to be in the range of  $-0.8$ to $-1.05$. Spectral index of NGC 2997 
which is a Sc galaxy lies in the above range. The flatter spectrum $\alpha=-0.79\pm0.07$ 
for the GMRT data at the lower frequencies compared to the wide band spectral index value 
could suggest a break in the spectrum at $\nu < 1\,\rmn{GHz}$. However the change in 
$\alpha$ is only $\sim2\sigma$ and needs to be confirmed with data at still lower $\nu$.\\
\hspace*{6mm}\citet{b6} estimated a global thermal fraction for most of the normal galaxies, 
as : $\rmn{S/S_{th}}\,\sim 1 + 10\left(\frac{\nu}{\rmn{GHz}}\right)^{(0.1-|\alpha_{nt}|)}$, 
where S is the total flux density, and $\rmn{S_{th}}$ is the flux density due to the thermal component. 
Taking, $\alpha_{nt}=-0.79$, thermal fractions derived at 1272, 616 and 332 $\rmn{MHz}$ are 
$\sim$ 10, 6 and 4 per cent respectively of the measured total flux density value at each 
observing band. The derived thermal fraction at 1.2 GHz agrees with the obtained mean 
value of thermal fraction at $1\,\rmn{GHz}$ for 74 galaxies by \citet{b45} which is 
$\langle\,\rmn{f}^{1\rmn{GHz}}_{th}\rangle=0.08\,\pm0.01$. The average thermal fraction 
at 1.4 $\rmn{GHz}$ in normal galaxies is $\lesssim 10$ per cent \citep{b21,b6}.\\
\hspace*{6mm}The standard method applied here to derive the thermal fraction
using the constant non-thermal spectral index is simplistic. \citet{b46} derived the 
thermal radio continuum in the galaxy M33 from the extinction corrected H$_\alpha$ map 
instead of using a constant $\alpha_{nt}$ across the galaxy. They find that the thermal 
fraction of M33 at $3.6$ cm using the new method is $23\,\pm14$ per cent lower than 
found by assuming a constant $\alpha_{nt}$ across the galaxy. However, \citet{b46} note 
that the thermal fraction derived from the radio integrated spectrum of M33 using
the standard method and the new method agree. Hence, assumption of a constant 
non-thermal spectral index to derive the thermal fraction is reasonable when the integrated 
spectrum is used.\\
\hspace*{6mm}Considering the small value of the thermal fraction in the total emission of 
NGC 2997 at low frequencies, we assume that the spectral power given in Table \ref{n2997gb} is 
mostly non-thermal. 
\normalsize
We estimate the supernova rate (yr$^{-1}$) in NGC 2997 from the non-thermal
luminosity \citep{b6}. The estimated supernova rate for the disc and nuclear region is about 
$\sim 0.08$ and $\sim 0.01$ $\rmn{yr}^{-1}$ respectively. 
In order to compare the SN rate per kpc$^2$, we assumed a diameter $\sim\,1.3$ 
and $\sim 24.5$ kpc for the nucleus and the disc respectively.
The SN rate per kpc$^2$ for the nucleus and the whole disc is about $\sim 8\,\times\,10^{-3}\,\rmn{yr}^{-1}\,\rmn{kpc}^{-2}$ and 
$\sim 0.14\,\times\,10^{-3}\,\rmn{yr}^{-1}\,\rmn{kpc}^{-2}$ respectively. 
If we assume a filling factor of $\sim 0.7$ for the disc, the SN rate for 
the disc is about $\sim 0.2\,\times\,10^{-3}\,\rmn{yr}^{-1}\,\rmn{kpc}^{-2}$ 
i.e. the nuclear region has a SN rate which is 40 times higher than the rate 
in the disc.\\
\hspace*{6mm}The tight correlation between the radio continuum 
intensities and the FIR luminosities is known to be linear 
over a wide range in star formation rate from normal spirals to the most intense starbursts
\citep{b47}. \citet{b49} found that Seyfert galaxies with radio quiet cores follow the linear 
radio-FIR correlation similar to normal galaxies. NGC 2997 is a 
late type spiral galaxy and having no obvious companion \citep{b26}. 
With the resultant steep spectral index $\alpha = -0.92\pm0.04$ of NGC 2997, 
absence of an active nucleus (see section 4.3), and the observed logarithmic 
measure of FIR/radio flux density ratio at our observing frequencies,
the linear correlation between FIR and radio emission exists in NGC 2997. 
For estimating various parameters like average formation 
rate of stars more massive than $5\,\rmn{M_{\odot}}$ 
(SFR($M\,\gtrsim 5 M_{\odot}$) $\rmn{M_{\odot}\,yr^{-1}}$) and the ionization rate of 
the Lyman continuum photons ($s^{-1}$) given in Table \ref{derived_par}, 
we assumed a simple model with only one free parameter $-$ the average formation 
rate of stars more massive than $5\,\rmn{M_{\odot}}$ \citep{b4}.\\
\begin{table}\caption{Total flux density measurements of NGC 2997}\label{tp_tab}
\scriptsize
\begin{tabular}{@{}lcc}
\hline
 Frequency  &   Flux density  &   Reference \\
  $\rmn{GHz}$       &     Jy          &               \\
\hline
  8.460     &  0.034 $\pm$0.004 &  \citet{b17}$^{a}$\\
  5.010     &  0.092 $\pm$0.010 &  \citet{b36}\\
  4.860     &  0.067 $\pm$0.011 &  \citet{b17}$^{a}$\\
  4.850     &  0.141 $\pm$0.014 &  \citet{b37}\\
  2.373     &  0.160 $\pm$0.010 &  \citet{b17}\\
  1.543     &  0.255 $\pm$0.020 &  \citet{b17}\\
  1.490     &  0.290            &  \citet{b3}\\
  1.400     &  0.235 $\pm$0.010 &  \citet{b7}\\
  1.272     &  0.367 $\pm$0.008 &  This paper\\
  0.616     &  0.730 $\pm$0.022 &  This paper\\
  0.332     &  1.134 $\pm$0.111 &  This paper\\
\hline
\end{tabular}\\
\footnotesize{a.Lower limits, because of missing spacing problems, not included in fit in Fig.\ref{tpcurv_fig}} 
\end{table}
\normalsize\hspace*{6mm}Radio counts of young Galactic SNRs suggest a radio supernova 
rate of 0.013 yr$^{-1}$ \citep{b2} or Type II supernova rate $\nu_{SN}\,\sim 0.023\,\rmn{yr^{-1}}$ 
estimated by \citet{b32}. Our data suggest that the supernova rate in NGC 2997 
is a factor of three to four times higher than that of our Galaxy. The Lyman 
continuum photon rate of the Galaxy is $2\times10^{53}$ photons s$^{-1}$ \citep{b16}. 
The derived star formation rate in NGC 2997 suggest that the galaxy is a normal galaxy 
with intermediate star formation. The derived values of star formation rate 
and the Lyman continuum photon rate of NGC 2997 are less than the values derived 
for M82 by \citet{b4} but higher than our Galaxy. For M82, the star formation 
rate SFR($M\gtrsim5M_\odot$) is $\sim2.2\,\rmn{M_{\odot}\,yr^{-1}}$, 
the ionization rate is $\rmn{N_{uv}}\sim8\times10^{53}\,\rmn{s^{-1}}$ and the radio 
supernova rate is $\nu_{SN}\sim0.1\,\rmn{yr^{-1}}$ \citep{b4}.
\begin{table}\caption{Estimated parameters for NGC 2997}\label{derived_par}
\scriptsize
\begin{tabular}{ l c c }
\hline
         &     Disc &  Nuclear region    \\
\hline
1. Star-formation rate ($\rmn{M_{\odot}yr^{-1}}$) & 1.9 & 0.2 \\
2. The ionization rate ($10^{53}\,$s$^{-1}$)  & 6.7& 0.8\\
3. The supernova rate (yr$^{-1}$) & 0.08  & 0.01 \\
\hline
\end{tabular}
\end{table}\normalsize~
\subsection{Spectral index (SI) distribution}

\begin{figure*}
\vspace*{2mm}
\begin{tabular}{@{}cc@{}}
\includegraphics[height=6.63cm,width=8.00cm]{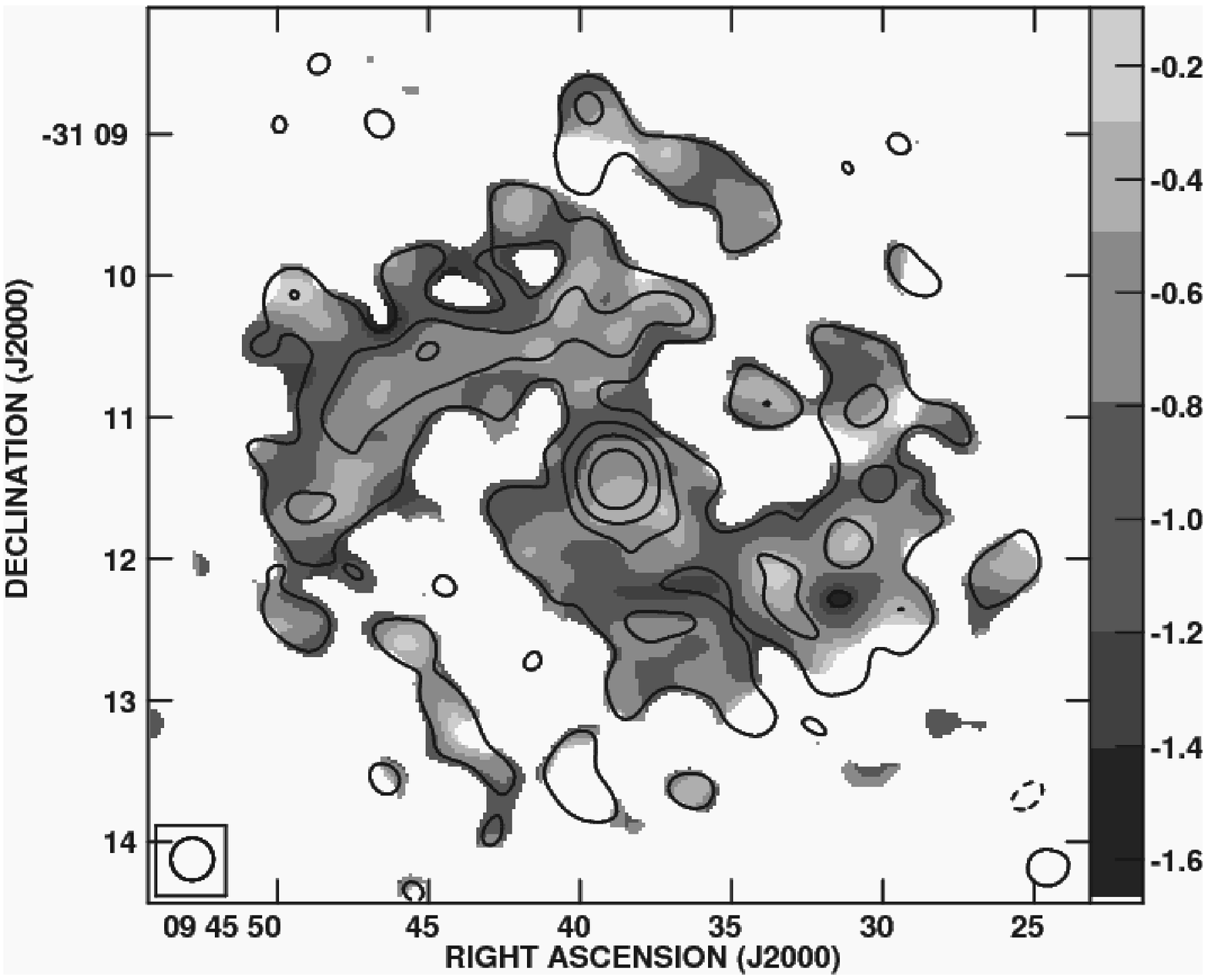} & \includegraphics[height=6.63cm,width=8.0cm]{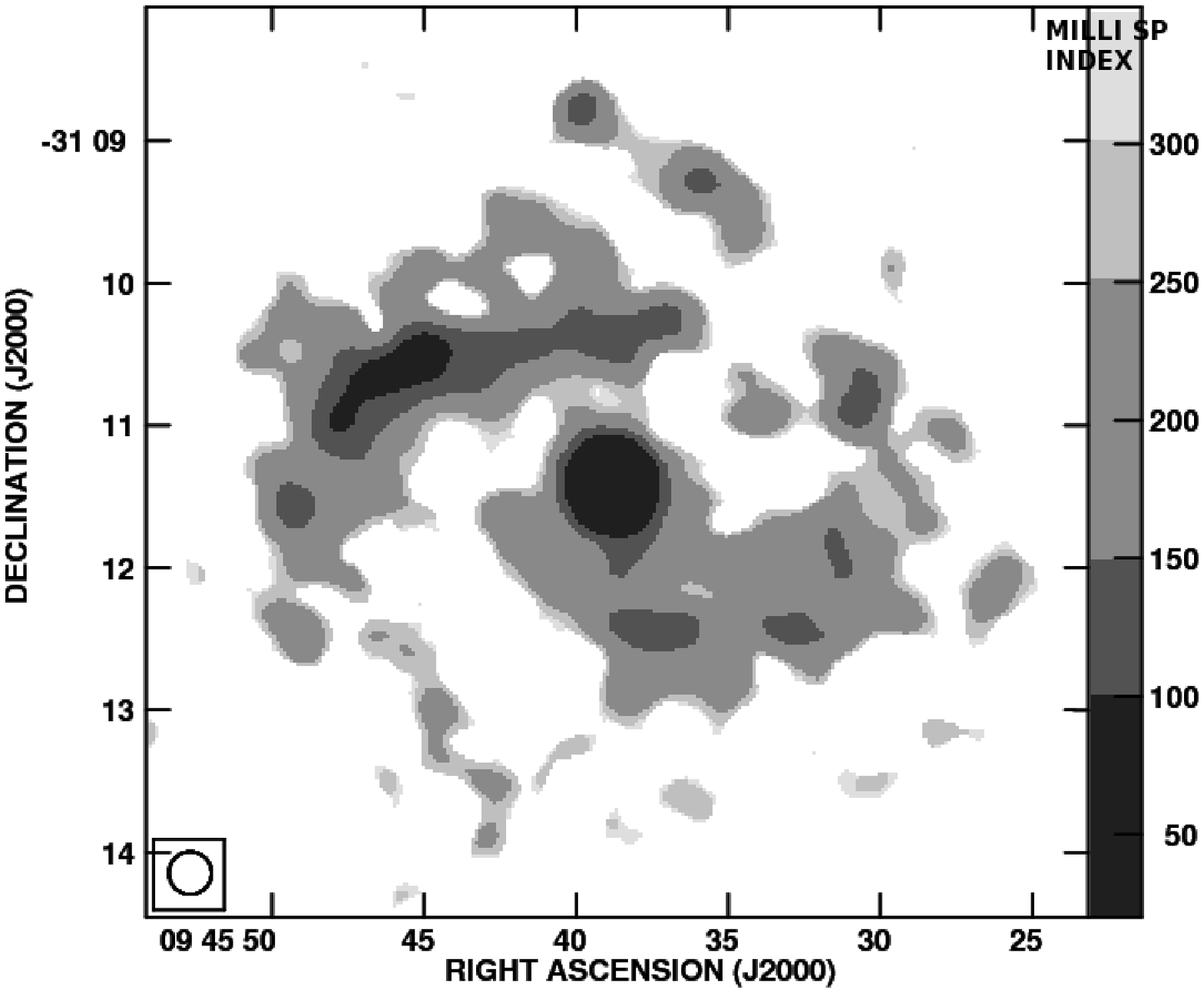}\\
\end{tabular}\caption{Left panel shows grey-scale spectral index image ($\alpha_{1.27,0.33}$) between 1272 and 332 $\rmn{MHz}$ and right panel shows the image of spectral index uncertainty. The radio contour map at 1272 $\rmn{MHz}$ superimposed on the grey-scale spectral index image with contours (-3, 3, 6, 12, 24)$\times\sigma$ level. The rms $\sigma$ is 0.5 mJy per beam area and the nucleus is the brightest region of the galaxy. All images in this figure are restored with a beamsize of 18 arcsec, the beamsize is indicated at the BLC.}\label{n2997sp_map}
\end{figure*}

The spectral index ($S_{\nu}\propto\nu^{\alpha}$) distribution across the galaxy 
was determined by using total intensity maps of identical beamsize at 1272 and 
332 $\rmn{MHz}$ respectively. Both the maps were convolved with a circular 
Gaussian beam of 18 arcsec in size and all points below $3\sigma$ level were 
blanked out. Spectral index map ($\alpha_{1.27, 0.33}$) was then derived using 
the COMB task in AIPS. 
The SI map between 1272 and 332 $\rmn{MHz}$ is shown as a grey scale image 
in Fig.\ref{n2997sp_map} (left) with the 1272 $\rmn{MHz}$ radio contour map 
superimposed on it. The uncertainty for the measured SI map is shown in 
the right panel of Fig.\ref{n2997sp_map}. The spectral index measured over most 
of the disc shows uncertainty in the range of 0.05 to 0.25. The uncertainty
map (Fig.\ref{n2997sp_map} right) shows intense regions (i.e. high S/N) having 
error $\lesssim 0.15$ whereas uncertainty increases in the fainter parts 
of the galaxy. Strong features registered with position errors of the order of 
3 to 4 arcsec in total intensity maps can make larger changes in the spectral 
index measurements. The errors in RA and DEC in our total intensity maps 
are $\sim\,\pm0.22^{s}$ and $\sim\,\pm 2$ arcsec, respectively. \\
\normalsize~
\normalsize
\hspace*{6mm}The SI map in Fig. \ref{n2997sp_map} shows that the radio spectrum
is flatter in nature wherever brightness peaks such as intense spiral arms and nucleus exist,
whereas radio spectrum of the outer parts of spiral arms show steeper spectrum.
The radio emission from normal galaxies is known to be closely related to the 
population of young massive main sequence stars which ionizes the HII regions causing
the free free thermal radio emission. The supernova explosions of these young massive 
stars accelerate the relativistic electrons causing the non-thermal synchrotron emission. 
As the spectral indices of Galactic supernovae remnants (SNRs) are known to show a 
broad range ($-0.2 < \alpha < -0.8$) with a mean value of $\sim -0.5$ \citep{b64} and considering 
that the thermal fraction of the observed total flux densities at low frequencies is less 
than 10 per cent, the relative flatter spectral index regions in spiral arms could 
suggest direct contribution of non-thermal emission by SNRs.
The spectral index value which increases up to $\sim -1.2\,\pm0.2$ in the outer parts
of the spiral arms indicates the energy losses of the relativistic electrons while 
they diffuse away out of their place of origin in star forming regions.\\
\hspace*{6mm}We measured a spectral index $-0.6\,\pm0.04$ for the nucleus 
by a single power law fit to the flux densities listed in Table \ref{n2997dtab} (source 1).
This is consistent with the mean spectral index value of the nucleus in the SI map 
shown in Fig.\ref{n2997sp_map} ($\alpha_{1.27, 0.33} \sim -0.62 \pm0.1$).
A relatively flatter spectral index of the nuclear region might be due to a 
younger population of particles, due to the thermal contribution by ionized 
gas in massive-star clusters and SNR.\\
\hspace*{6mm}The spatial spectral index for the ridge-component in the northern 
spiral arm varies from $-0.78 \lesssim \alpha_{1.27, 0.33} \lesssim -0.46$. 
The source $3$ in the ridge-component (Fig.\ref{n2997sp_map}, left) 
is a giant HII region and the core source in this region is classified as a 
thermal source \citep{b12}. The spectral index value for source 3 is $\sim -0.06$ 
(see Table \ref{n2997dtab}) and confirms the thermal nature of emission.
\normalsize
\subsection{The circumnuclear ring}

\begin{figure*}
\begin{center}
\begin{tabular}{@{}cc@{}}
\includegraphics[height=6cm,width=7.2cm]{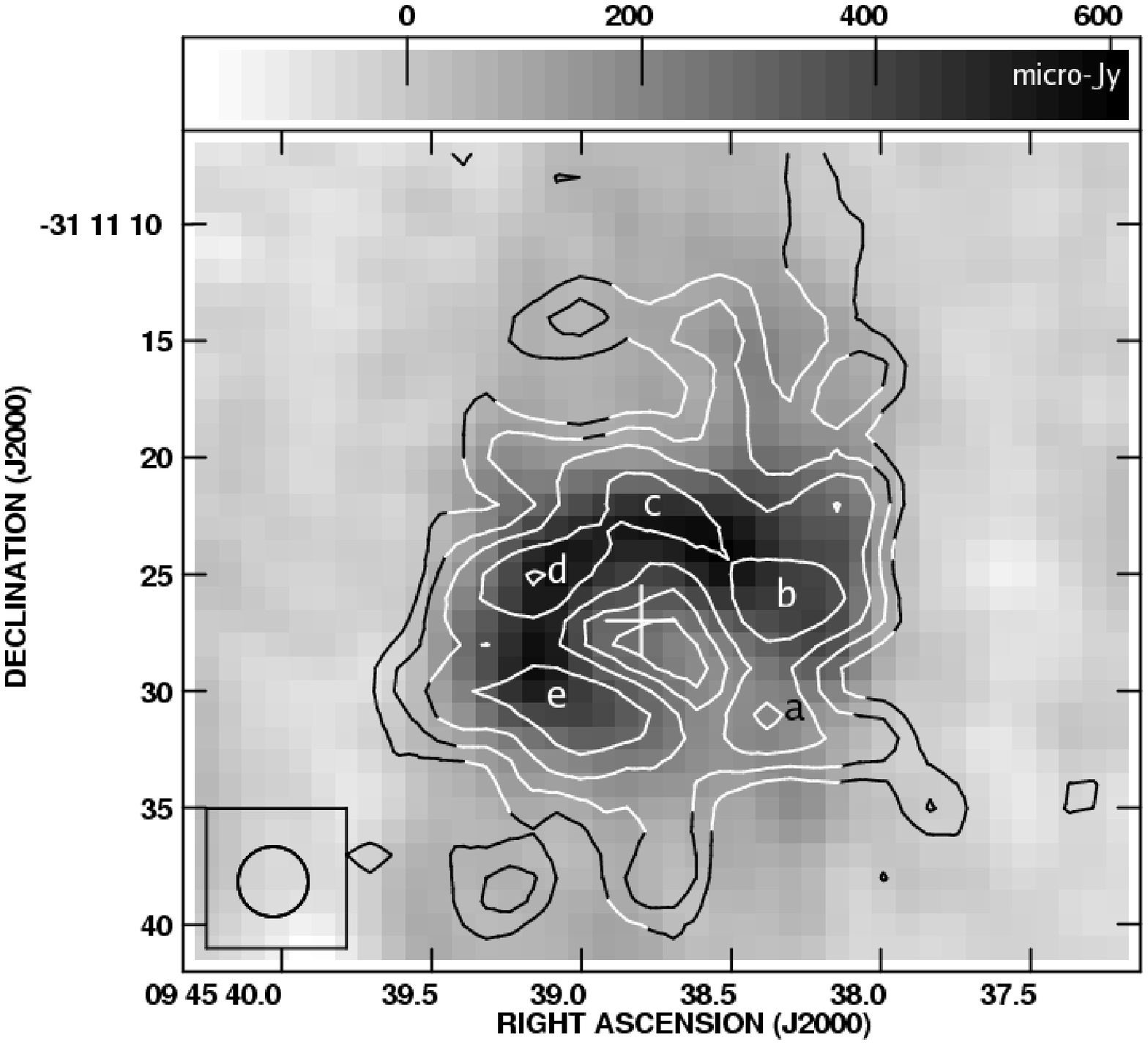} & \includegraphics[height=5.6cm,width=7.2cm]{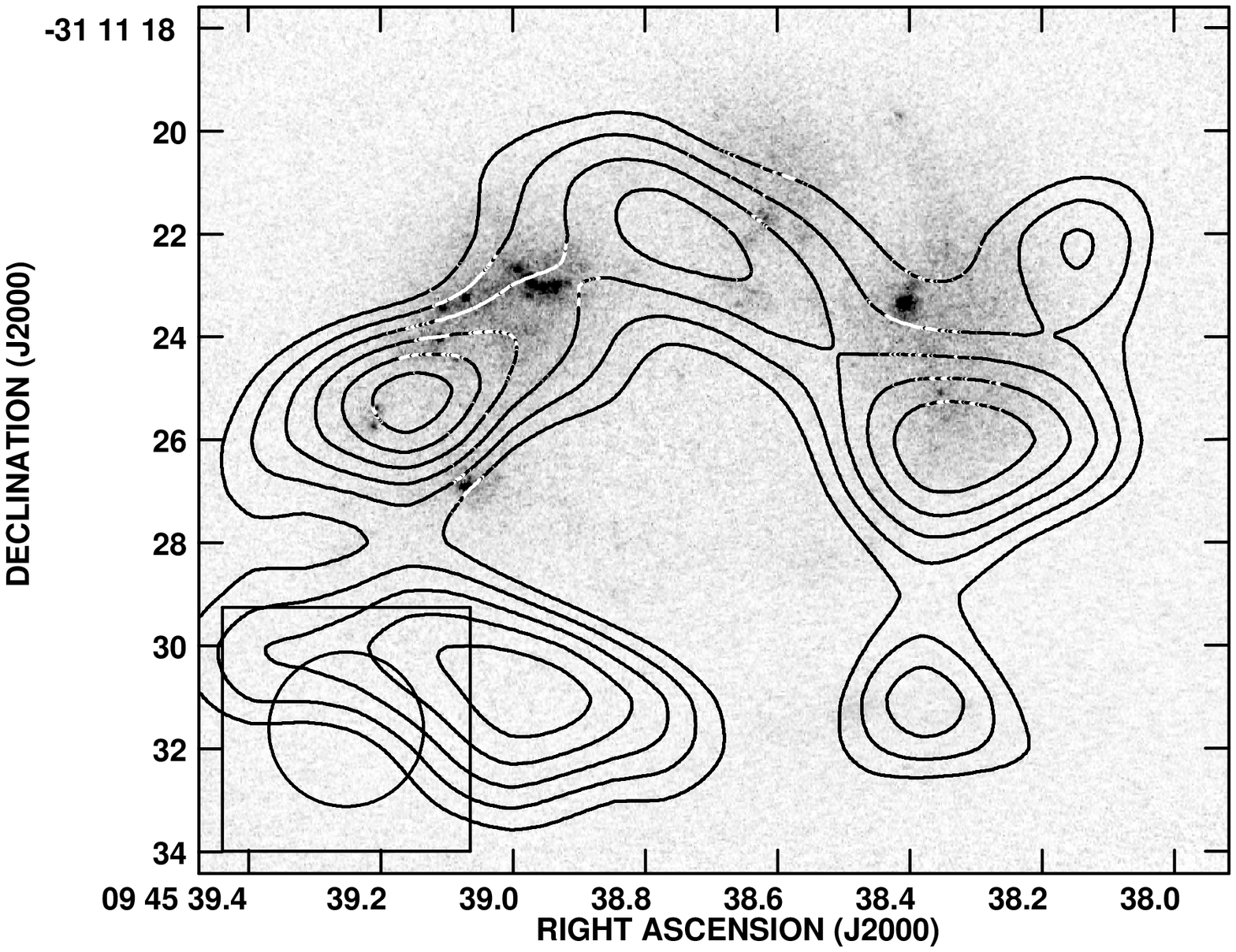}\\
\parbox[t][8mm]{65mm}{\footnotesize{\textbf{\small{Fig.5 left:}} Contour levels are at 0.24, 0.4, 0.56, 0.72, 0.88, 1.04, 1.2, 1.36 $\rmn{mJy\,beam^{-1}}$, the peak brightness is 1.39 $\rmn{mJy\,beam^{-1}}$}} & \parbox[t][8mm]{65mm}{\footnotesize{\textbf{\small{Fig.5 right:}} Contour levels are at 0.72, 0.84, 0.96, 1.07, 1.2, 1.32 $\rmn{mJy\,beam^{-1}}$, the peak brightness is 1.39 $\rmn{mJy\,beam^{-1}}$}}\\
\end{tabular}
\caption{\small{The GMRT radio contour map of the nuclear region at 1.27 $\rmn{GHz}$ with an angular resolution of $3\times3$ arcsec$^2$ overlaid on (left) gray-scale archival VLA 4.8 $\rmn{GHz}$ {\it C\/} band image ( restored with a beamsize $3\times3$ arcsec$^2$, PA$=0^{\circ}$) and (right) gray-scale {\it HST\/} archival {\it V\/} band $2300 \rmn{\AA}$ image (Courtesy, \citet{b9}; resolution $\sim 0.05$ arcsec,PA$=84.30^\circ$).} }\label{circumn2997fig}
\end{center}
\end{figure*}
We detect a strong nuclear source in NGC 2997 at all the wavebands which is
resolved into a circumnuclear starforming ring in our high resolution
1272 MHz image. NGC 2997 is an intermediate type SAB(rs)c galaxy with 
several hotspots. The circumnuclear star forming ring has been observed 
in the optical and NIR \citep{b9,b11}. 
Circumnuclear rings are observed in 20 per cent of all spiral galaxies 
and mostly occur in barred galaxies \citep{b50}. Observational studies \citep{b56,b55} 
show that the circumnuclear ring can arise due to the bar driven radial inflow of 
magnetized gas materials which gets accumulated in a ring near the location of inner 
Lindbland resonances. There are two models which explain the star formation
in circumnuclear rings : (i) in the gravitational instability or ``popcorn" \citep{b57,b55} model,
star formation is driven by stochastic gravitational fragmentation 
along the ring where the star forming regions have a regularly spaced distribution \citep{b11}. 
(ii) in the ``pearls on a string" model, the gas along the bar that
flows into the ring is compressed near the contact points. Star formation is then 
triggered in these over density contact regions. Observationally an age gradient
in the star forming regions is seen along either half of the ring \citep{b56,b55}.\\
\normalsize
\hspace*{6mm} We made a high resolution image at 1.27 $\rmn{GHz}$ using robust 
weighting of $-5$ in the task IMAGR. The resulting angular resolution of 3 arcsec 
($\sim$ 200 pc) resolved the circumnuclear ring (see Fig.\ref{circumn2997fig}) 
into five star-forming clumps i.e. hotspots. We measured the position, peak intensity 
and integrated flux-density of each hotspot by fitting a Gaussian using the task JMFIT in AIPS. 
The position for each component is listed using seconds of RA and arcsec of DEC in Table 
\ref{circumn2997tab}. The observed luminosity for the individual hotspot is between 
0.16 to 1.58
 $\times\,10^{20}\,\rmn{W\,Hz^{-1}}$ at $20$ cm band with a $200$ pc resolution. 
We also calibrated and imaged VLA archival data of this galaxy at 
4.8 $\rmn{GHz}$. The resulting image shown as grey scale in Fig.\ref{circumn2997fig} 
(left) also detects the circumnuclear ring coincident with the 1.27 $\rmn{GHz}$ ring. 
Fig.\ref{circumn2997fig} (right) shows the 1.27 $\rmn{GHz}$ radio contour map 
overlaid on {\it HST\/} {\it V\/} band gray-scale image \citep{b9}. The 
radio ring is coincident but more extensive than the observed structure of the 
ring in the {\it V\/} band. This is likely to be due to the lower angular resolution at 
radio wavelengths and partly probably due to extinction at optical wavelengths which 
\citet{b11} estimate to be about 3 magnitudes in the vicinity of the ring. The 
morphological structure of the circumnuclear region in our 1.27 $\rmn{GHz}$ radio 
contour map is similar to the contour maps of relative intensities in {\it U\/} 
band published by \citet{b24} using the SAAO 1.9-m reflector.\\
\hspace*{6mm}Our radio contour map and the gray-scale images in the {\it UV\/} ($2300\,\rmn{\AA}$) 
and radio (4.8 $\rmn{GHz}$) band shown in Fig.\ref{circumn2997fig} indicate that 
there is no detectable radio continuum emission from the central part of the nucleus,
confirming that there is no radio-loud AGN in the center of this galaxy. 
\citet{b9} observed circumnuclear rings in five barred or weakly barred spiral galaxies 
of type Sc (including NGC 2997) and none of the rings hosted an AGN at its centre.
The quiescent phase of the nucleus was also noted in the detailed study of the NGC 2997 
nucleus by \citet{b35}. 
A large number of hotspots are observed on a complete circumnuclear ring
of NGC 2997 at NIR \citep{b11} whereas only three northern star-forming regions
are visible in the {\it V\/} band image. We note that the 
intense NIR hotspots located in the northern part of the ring correlate with 
the radio hotspots. Weak radio emission is detected from the southern parts 
of the ring although hotspots are not distinguishable. The faint emission along 
the southern half of the ring is seen in {\it UV\/} \citep{b9}, {\it U\/} \citep{b24} 
band image. Faint southern hotspots are also seen in NIR ring which 
\citet{b11} attribute to the heavy dust extinction since the southern hotspots are being
observed through the dust lanes. A slight offset between the optical and radio positions 
of the hotspots are observed. This could be attributed to the poor angular resolution 
of the radio data ($\sim 3$ arcsec) as compared to the excellent
resolution of the optical data ($\sim0.05$ arcsec) and dust 
extinction near the star forming peaks.\\
\hspace*{6mm}The circumnuclear ring that we detect in radio has a deconvolved diameter 
of $11.6\,\pm0.08$ arcsec i.e. $\sim 750$ pc. 
The separation between the hotspots ranges from $\sim\,190$ to $370$ pc. 
This is comparable to the NIR where the diameter of the ring is $\sim\,570$ pc ($\sim\,$8.7 arcsec) 
and the spacings between hotspots is $\sim\,200$ pc \citep{b11}.
This is typical of the circumnuclear rings detected in other galaxies. 
The average size of the nucleus from the photometric and morphological analysis of 
the galaxies with peculiar nuclei is quoted to be less than $800$ pc \citep{b28}.\\
\hspace*{6mm}We compared the circumnuclear ring (CNR) in NGC 2997
to the well studied CNR in NGC 1097 \citep{b35,b58,b54,b59}. The CNR of NGC 1097
has $\sim\,1.5$ kpc diameter \citep{b60} and a LINER/Seyfert 1 AGN at its centre.
Three prominent radio hotspots are seen at $3.5$ cm in NGC 1097 \citep{b60} whereas
four bright and one faint hotspots are observed in NGC 2997 at $20$ cm (see Fig.\ref{circumn2997fig}).
No azimuthal age gradient is seen in massive stars of both the rings with ages less 
than $10^7$ to $10^8$ yrs \citep{b9,b11,b59} and both rings show regular distribution 
of hotspots \citep{b58,b11,b60} which lends support to the gravitational instability model \citep{b57}
for the star formation in the CNR. The CNR in NGC 1097 is formed due to the gas driven by bar potential. 
Various observational studies show that NGC 2997 does not possess either a prominent 
large scale bar or an active nucleus \citep{b35,b9}, hence unlike NGC 1097 gas accumulation 
in the CNR of NGC 2997 is unclear.
However, Fast Magnetohydrodynamic Density Waves (FMDWs) at the modified inner Lindbland resonance 
(mILR) model \citep{b63,b62} can explain the formation of CNR in NGC 2997.\\
\hspace*{6mm}We estimate an equipartition field in the central nuclear region
of diameter $\sim\,750$ pc to be about $30\,\mu$G. This is similar to the field
strength of $34\,\mu$G estimated by \citet{b17} for the central region of NGC 2997
and $40\,\mu$G estimated for the CNR in NGC 1097 by \citet{b54}. The equipartion 
field for the radio hotspots in NGC 2997 ranges from $\sim 30$ to $50\,\mu$G.
We estimate a disc field of $\sim\,4\,\mu$G for NGC 2997.\\
\begin{table*}
\centering
\begin{minipage}{140mm}
\caption{Radio hotspots in the circumnuclear region of NGC 2997 }\label{circumn2997tab}
\scriptsize
\begin{tabular}{@{}cccccc@{}}
\hline\\[-2.0ex]
Hotspot & Position (J2000)        & Deconvolved angular & Peak Intensity & Integrated  & Radio luminosity \\
  & RA (sec), DEC (arcsec)& size (arcsec$^2$ PA) & ($\rmn{mJy\,beam^{-1}}$)    & flux density ($\rmn{mJy}$)& ($\,10^{20}\,\rmn{W\,Hz^{-1}}$)\\
\hline\\[-2.0ex]
a & 38.37,31.01 &  1.7  $\times$ 1.3 @167$^\circ$& 0.61 & 0.77 $\pm0.17$ & 0.16 \\
b & 38.32,25.77 &  7.5  $\times$ 5.0 @ 88$^\circ$& 1.34 & 7.0  $\pm0.5$& 1.52 \\
c & 38.78,22.31 & 12.5  $\times$ 4.0 @ 76$^\circ$& 0.77 & 5.0  $\pm0.6$& 1.17 \\
d & 39.12,25.00 &  7.7  $\times$ 2.1 @122$^\circ$& 0.96 & 3.27 $\pm0.3$& 0.7 \\
e & 39.02,30.86 &  9.4  $\times$ 4.0 @71 $^\circ$& 1.34 & 7.36 $\pm0.5$& 1.58 \\
\hline
\end{tabular}
\end{minipage}
\end{table*}
\hspace*{6mm}
We compared the luminosity of radio hotspots in circumnuclear region 
of NGC 2997 with the circumnuclear radio hotspots in other galaxies such as NGC 253, 1365 and 
1808. NGC 253 is a nearby ($\rmn{d}\sim 2.5$ Mpc) starburst galaxy, in which \citet{b40} detected 
22 compact circumnuclear sources at $20$ cm with $\sim 28$ pc resolution. The detected 
circumnuclear sources of NGC 253 have typical radio power of $2 \times 10^{18}\,\rmn{W\,Hz^{-1}}$. 
\citet{b41} detected seven hotspots in the circumnuclear region of NGC 1365 at 3 and $6$ cm 
with a $100$ pc resolution ($\rmn{d}\sim 20$ Mpc). The luminosity of circumnuclear hotspots 
for NGC 1365 at $20$ cm (derived using a SI from 6 and $3$ cm) ranges from 
$\sim 0.8\,\rmn{to}\,2\,\times 10^{20}\,\rmn{W\,Hz^{-1}}$. Similarly, \citet{b42} detected hotspots 
in NGC 1808 with a resolution of $98$ pc. The luminosity of circumnuclear hotspots in NGC 1808 
($\rmn{d}\sim 9.9$ Mpc) at $20$ cm (derived using a SI between 6 and $3.6$ cm) ranges 
from $\sim 0.2\,\rmn{to}\,3.8\,\times 10^{20}\,\rmn{W\,Hz^{-1}}$. 
Thus, the average luminosity of hotspots in a CNR observed with $\sim 100$ pc 
resolution appears to be $10^{19}$ to a few times $10^{20}\,\rmn{W\,Hz^{-1}}$ 
at $20$ cm. It is likely that the hotspots are composed of several 
$2\,\times 10^{18}\,\rmn{W\,Hz^{-1}}$ clumps as seen in NGC $253$ in addition 
to diffuse non-thermal emission. Assuming that the observed emission at 1.27 $\rmn{GHz}$ 
from the circumnuclear hotspots is mostly non-thermal in nature, we derived the SN 
rate based on \citet{b6} equation. The median SN rate value for the discrete 
components in NGC 2997 ring is $\sim 0.001$ yr$^{-1}$. The median value of 
average formation rate of stars (SFR($M \ge 5 M_{\odot}$) M$_\odot$ yr$^{-1}$) 
estimated for the hotspots is $\sim 0.024\,$M$_\odot$ yr$^{-1}$.\\
\hspace*{6mm}The spectral index $\alpha_{4.8,1.27}$ of the CNR in NGC 2997 
is $\sim-0.64\,\pm0.1$. We could not estimate the 
spectral index of the individual hotspots since these were not clearly 
discernible in the convolved 4.8 GHz map (see Fig.\ref{circumn2997fig}). 
NGC 6951, unlike NGC 2997, hosts a central AGN and has a marginally 
larger diameter ($\sim 1$ kpc) of the CNR but both the rings show hotspots 
which form an almost complete ring. In NGC 2997, the radio ring is fainter 
to the south whereas in NGC 6951 the ring appears to be broken
 in the north 
and south \citep{b30}. The spectral index of the circumnuclear emission 
in NGC 6951 between 20 and $6$ cm was found to be $-0.84$ \citep{b43}. 
The radio spectrum of the hotspots in the CNR of other galaxies 
exhibit a large range in spectral index. For example, the 
CNR in NGC 613 has a spectral index of -0.65 between 2 and $6$ cm 
\citep{b19} similar to NGC 2997, whereas the spectral index of 
the individual compact sources (total 19) in the CNR of NGC 4736 
exhibit spectral index ranging from 0.1 to $-0.8$ between $6$ and $20$ cm \citep{b10}. 
Similar kind of distribution is also noted in the CNR of NGC 1365 between $3$ and $6$ cm 
by \citet{b41}. Out of seven hotspots detected in the ring of NGC 1365, 
four hotspots show steep spectrum and rest show flat spectrum with 
the spectral index ranging from 0.04 to $-0.4$ \citep{b41}. \citet{b41} attribute 
synchrotron emission from hotspots to electrons accelerated by SN 
and SNRs. The spectral indices of circumnuclear hotspots detected between $3.6$ 
and $6$ cm in NGC 1808 \citep{b42} are also consistent with them being dominated by SNRs. 
Utilizing a {\it V\/} band {\it HST\/} exposure of NGC 2997, \citet{b9} determined 
the numerous compact (few pc radius) sources distributed along the ring that are 
probably young (less than $10^8$ yr) and massive ($\sim10^5$ M$_\odot$) clusters of 
stars that are gravitationally bound.
\section{Summary}

We have presented high resolution radio continuum maps of NGC 2997 at 332, 616 and 
1272 $\rmn{MHz}$. To the best of our knowledge these are the first images of the galaxy
at $\nu < 1.4$ $\rmn{GHz}$. The radio continuum features detected in our observation
bands are described, in detail, along with the spectral index values. The best-fitting 
single power law model to the integrated radio spectrum of NGC 2997 gives a spectral 
index $\alpha = -0.92\pm0.04$. The best-fitting value of $\alpha = -0.79\pm0.07$ obtained 
using only the low-frequency GMRT data suggests a break in the spectrum at 
$\nu < 1$ $\rmn{GHz}$. We put an upper limit of 10 per cent on the thermal fraction 
at $1\,\rmn{GHz}$.\\
\hspace*{6mm}The spectral power of NGC 2997 at our observing frequencies is in 
the range of 7.9 to 24.4$\,\times\,10^{21}\,\rmn{W\,Hz^{-1}}$. The derived logarithmic 
measure of FIR/radio flux density ratio, $\hat{q}_{\rmn{FIR}}$ parameter is $\sim 2.19 \pm0.03$, 
which is near the median for spiral galaxies at 1.4 $\rmn{GHz}$, $\hat{q}_{\rmn{FIR}}\approx 2.30$ \citep{b4}.
The estimated values for average star formation rate SFR($M\gtrsim 5M_{\odot}$) 
is 1.9 M$_\odot$ yr$^{-1}$, the supernovae rate is 0.08 $\rmn{yr^{-1}}$, 
and the production rate of the Lyman continuum photons is $6.7\,\times\,10^{53}\,\rmn{s^{-1}}$. 
The resultant logarithmic measure of FIR/radio flux density ratio and the intermediate 
average star formation rate confirms that NGC 2997 is a normal galaxy.\\
\hspace*{6mm}The low frequency spectral index distribution map ($\alpha_{1.27,0.33}$) 
of NGC 2997 shows that the radio spectrum is flatter in nature near brightness 
peaks such as in intense spiral arm regions and nucleus, whereas radio emission 
from the low surface brightness regions shows a steeper spectrum. The measured 
spectral index for the nucleus between 332 and 1272 $\rmn{MHz}$ is $\alpha \sim -0.6 \pm0.04$.\\
\hspace*{6mm}The radio continuum map at 1272 $\rmn{MHz}$ with an angular resolution 
of 3 arcsec ($\sim 200$ pc) reveals a circumnuclear ring structure with five radio 
hotspots. No radio continuum emission at the centre of the galaxy is detected 
indicating that there is no radio-loud AGN. Part of the ring to the south is faint. 
The circumnuclear ring is similar to that observed at other wavebands from the galaxy 
and in other galaxies. It has a deconvolved diameter of $\sim 750$ pc and the 
separation between hotspots range from $\sim\,190$ to $370$ pc. We estimate an 
equipartition field in the central nuclear region of diameter $\sim 750$ pc
to be about $30\,\mu$G. The observed luminosity for the individual hotspot 
is about $10^{20}\,\rmn{W\,Hz}^{-1}$. The median SN rate and average formation 
rate of stars SFR($M\gtrsim 5M_{\odot}$) for the hotspots in the ring are 
$0.001\,\rmn{yr^{-1}}$ and $0.024\,\rmn{M_{\odot}\,yr^{-1}}$ respectively.
\footnotesize
\section*{Acknowledgments}

We thank the staff of the GMRT who have made these observations possible. GMRT is run by the
National Centre for Radio Astrophysics of the Tata Institute of Fundamental Research.
We thank the anonymous referee for useful comments on the manuscript. J. Kodilkar thanks the 
Dean of the GMRT, Prof. Y. Gupta for his support. One of the authors (SA) wishes 
to thank R. Wielebinski for bringing to his notice the importance of low frequency observation 
of NGC 2997. This research has made use of the NASA/IPAC Extragalactic Database which is operated by 
the Jet Propulsion Laboratory, Caltech under contract with the NASA.

\label{lastpage}
\end{document}